\documentclass[11pt]{amsart}
\usepackage{geometry}                
\usepackage{color}   
\usepackage{graphicx}
\usepackage{amssymb}
\usepackage{epstopdf}
\usepackage{hyperref}
\title{Counting muons to probe the neutrino mass spectrum}
\author[Carolina Lujan-Peschard, Giulia Pagliaroli, Francesco Vissani]{Carolina Lujan-Peschard$^{1,2}$, Giulia Pagliaroli$^{1}$, Francesco Vissani$^{1,3}$\\
$^{1}$ \tiny INFN, Laboratori Nazionali del Gran Sasso, Assergi (AQ), Italy\\
$^{2}$ Departamento de Fisica, DCeI, Universidad de Guanajuato, Le\'{o}n, 
Guanajuato, M\'{e}xico\\
$^{3}$ Gran Sasso Science Institute (INFN), L'Aquila, Italy}

\begin{document}
\maketitle

\begin{abstract}
The experimental evidence that $\theta_{13}$ is large opens new opportunities to identify  
the neutrino mass spectrum. We outline a possibility to investigate this issue by means of conventional technology. The ideal setup turns out to be long baseline experiment:  
the muon neutrino beam, with $10^{20}$ protons on target, has 
an average energy of 6 (8) GeV; the neutrinos,
after propagating  6000 (8000) km,  are  observed 
 by a muon detector of 1 Mton and with a muon energy threshold of 2 GeV.   
 The expected number of muon events is about 1000, and the 
difference between the two neutrino spectra  is sizeable, about 30\%. 
This allows the identification of the mass spectrum just counting muon tracks.
The signal events are well characterized experimentally by their time and direction of arrival,  and 2/3 of them are in a region with little atmospheric neutrino background, namely, between 4 GeV and 10 GeV. 
The distances from  CERN to Baikal Lake and 
from Fermilab to KM3NET, or ANTARES, fit in the ideal range.

\end{abstract}

\parskip3mm
\section{Introduction}

\begin{figure}[t]
\centerline{\includegraphics[width=13cm]{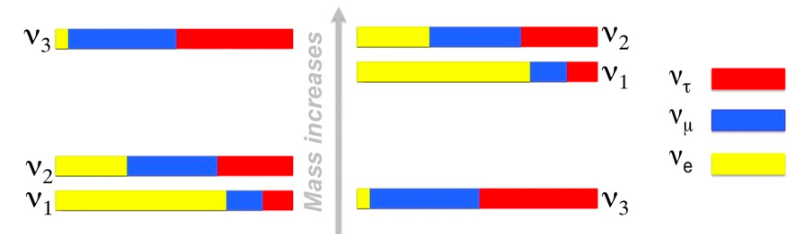}}
\caption{\footnotesize The two spectra that are compatible with present neutrino data: left, normal hierarchy; right, inverted hierarchy.  The flavor content of each individual state with
given mass, $|U_{\ell i}^2|$, is represented using the color code given in the rightmost part of the figure. \label{fig0}}
\end{figure}

The study of neutrinos from the sun \cite{ooosun}, from the Earth's atmosphere \cite{oooatm}
and from artificial sources \cite{oooart} led to discover neutrino
oscillations \cite{ooo} proving the  relevance of the matter effect
\cite{msw,msw1} for solar neutrinos. 
The most direct extension of the standard model requires 3 massive neutrinos. Such a picture, 
summarized following \cite{smirnovold}  in Fig.~\ref{fig0}, permits us
to account for most neutrino observations and to ask new questions;  in particular, 
how to probe whether the neutrino spectrum is normal or inverted. 
After the experimental evidence that $\theta_{13}$ is large \cite{meas13} 
a more precise formulation of this question is: how to take advantage
of the matter effect in the Earth to solve this ambiguity.

Note that the experimental approaches that are presently implemented (or those  planned  
before  $\theta_{13}$ was known) have not been 
optimized to observe a large matter effect, and seem 
unable to guarantee  very clear answers on which is the mass spectrum: 
see the recent  {\tt nuTURN} conference \cite{nuturn}, 
where this issue has been debated
and the conclusions of \cite{poco}.
However, the most relevant energies are indicated directly by the theory:
the matter effect is maximum when the matter and the vacuum terms, that describe oscillations, 
are comparable; this happens when the neutrino energies are in the range  $5-10$ GeV. 
Moreover, in order to have a large matter effect in the Earth, 
neutrinos have to cross a sizeable amount of matter,  
of the order of the Earth's radius, as discussed later on (Sect.~\ref{mp}) and in
agreement with the `oscillograms' of \cite{smirnov} or also with Fig.~3 of \cite{lathuile}.

An idea to proceed is to try to identify an experimental sample of atmospheric neutrino events where the matter effect is large; in order to achieve this goal, muon neutrinos and antineutrinos 
have been considered, because the muons and the antimuons are supposed to be   
identifiable experimentally. Several specific options have been considered: e.g., a 50 kton magnetized iron detector, able to measure the charge of the muon  \cite{mino} and to distinguish between neutrinos and antineutrinos;  an argon   detector of many 100 kton, able to measure also the energy of  the hadrons scattered by the neutrinos, reconstructing better the neutrino  energy \cite{gb}; a huge, $\sim 10$ Mton water Cherenkov muon detector \cite{smirnov}, as a dense core of underwater/underice installations aimed at seeing high-energy neutrinos from cosmic sources.
In any case, a very large detector mass is required, due to the need to probe relatively high energies, where the atmospheric neutrino flux is low.
Furthermore, the use of atmospheric neutrinos implies certain limitations: 
1) It is not possible to reconstruct the neutrino energies 
simply observing the muons, as in the largest water Cherenkov 
detectors, {because} a muon neutrino with
energy $E$  will give a muon with energies from zero to this value. 
2)~If muons and antimuons cannot be distinguished, the impact of matter effect is 
diluted. 

With these considerations in mind, we would like to suggest a different 
approach to emphasize the matter effect, 
insisting on the use of large (but relatively simple) muon detectors. 
We propose to send a muon neutrino beam produced in
laboratory, 
{maximizing the {\em muon neutrino disappearance} that is induced by the matter effect}
by a suitable choice of  the distance of propagation and of the average energy of the 
beam.\footnote{{Ref.\ \cite{wint,sp}  
have discussed electron neutrino appearance; however, 
the detection of a signal due to electrons is significantly more demanding than those due to muons.
Ref.~\cite{dick} discussed something closer to the present proposal; however, its authors have considered energy thresholds modeled on the existing neutrino telescopes, that are larger than our threshold: this has a significant impact on the ensuing conclusions.}}
{In the next section, 
we show that it is possible to arrange for a considerable amount of muon disappearance.} 
Based on this remark, we argue that 
it is possible to achieve the 
goal of measuring the neutrino mass hierarchy with conventional technology, and moreover, 
by employing a  well-known type of neutrino beam and possibly taking advantage of  sites 
that already host muon detectors. However, we need to improve 
the existing detectors, lowering their 
energy threshold to a few GeV. 
In short, we argue that in order to distinguish normal from
inverted hierarchy, we should simply count muons in the right type of experimental setup.

\section{\label{mp}Muon survival probability}
Let us begin the discussion 
from the case 
when only the largest $\Delta m^2$ matters. 
If $\theta_{13}$ is set to zero, we have 
2-flavor vacuum oscillation:
the muon survival probability 
oscillates as a function of the energy, its maxima correspond to the
minima of $P_{\mu\tau}$ (and viceversa), 
and there is no difference between normal and inverted hierarchy. 
Then, let us consider the effect of $\theta_{13}$.  
Assuming normal hierarchy, 
 $P_{\mu e}$ is amplified due to matter effect for certain energies. Therefore, around 
 these energies,  the muon survival probability,
\begin{equation}P_{\mu \mu}=1-P_{\mu e} - P_{\mu \tau}\end{equation}
must decrease. 
{Stated otherwise, we remark that the occurrence of electron neutrino appearance due to matter effect, i.e., an increase of $P_{\mu e}$, is strictly connected with the occurrence of muon neutrino disappearance, i.e., a decrease of $P_{\mu\mu}$, that is particularly remarkable 
when $P_{\mu\tau}$ is small.}
For inverted hierarchy, instead, $P_{\mu e}$ is suppressed by matter effect and the oscillations 
are very similar to the case when $\theta_{13}$ is set to zero.

\begin{figure}[tbp]
\begin{center}
\centerline{\includegraphics[width=7.4cm]{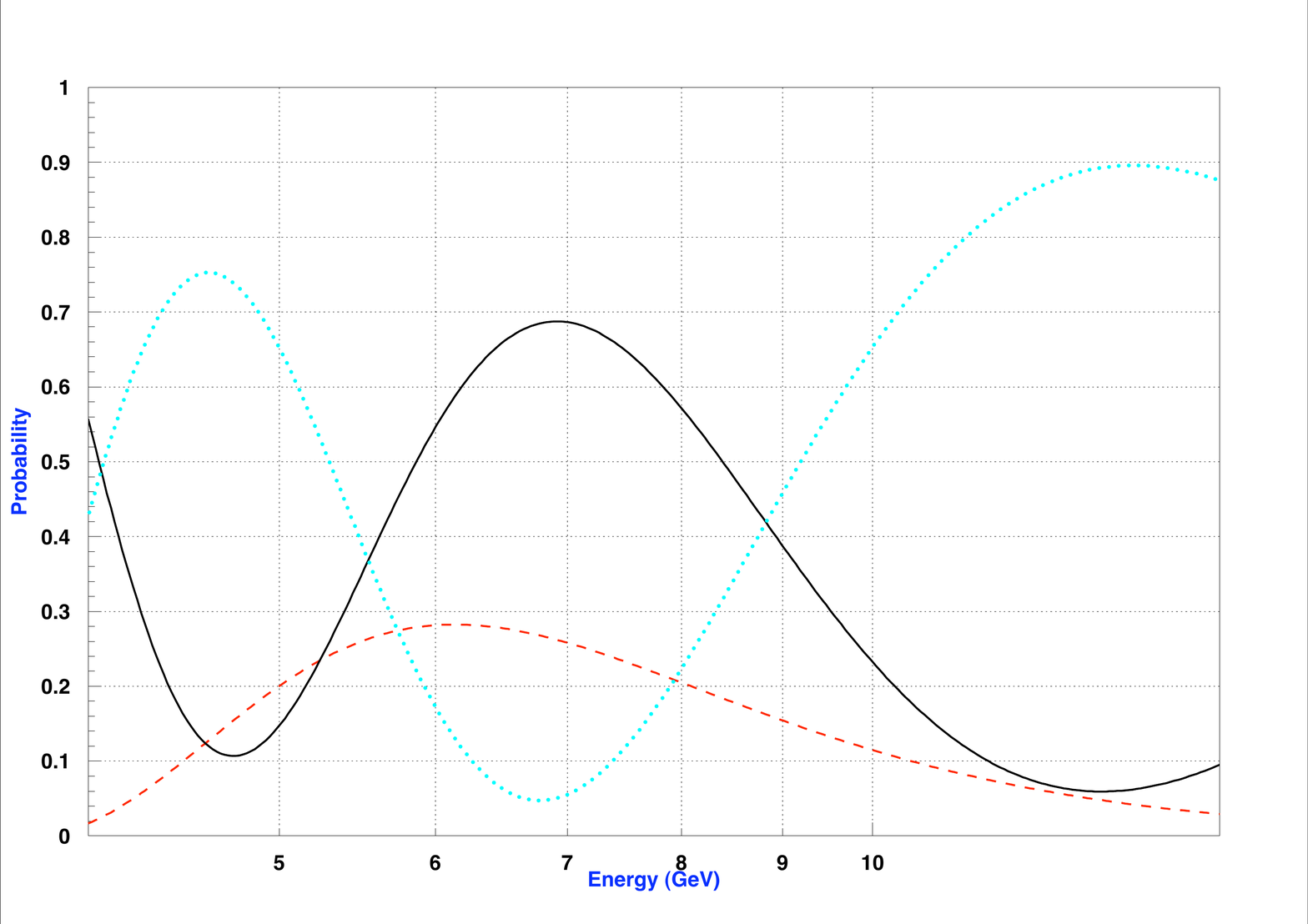}
\includegraphics[width=7.4cm]{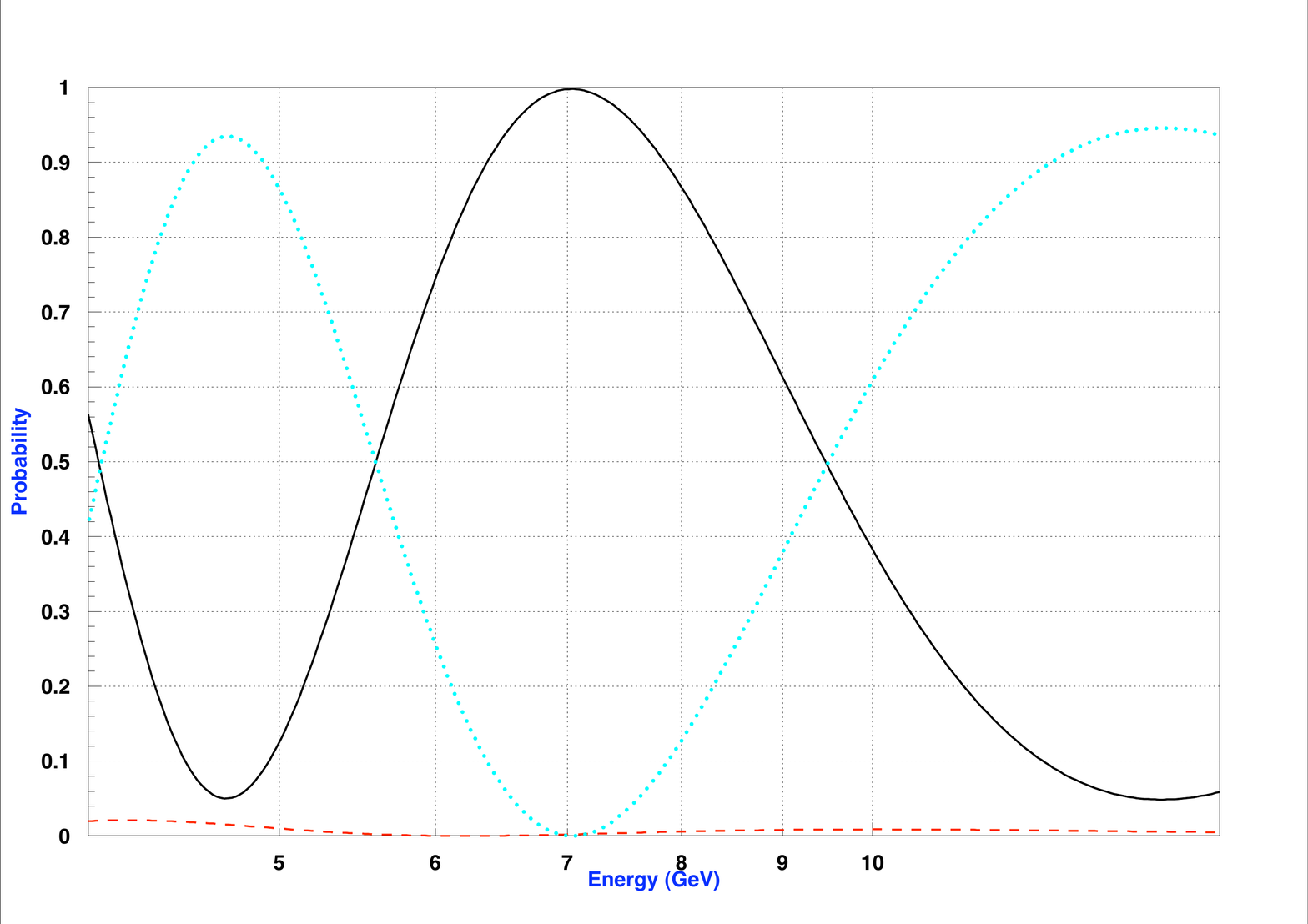}}
\caption{\footnotesize Oscillation probabilities for $L=7000$ km:
the continuous (black) line is $P_{\mu\mu}$; the dashed (red) line  $P_{\mu e}$;
the dotted (cyan) line $P_{\mu\tau}$. Left/right panel, normal/inverted hierarchy.
The increase of $P_{\mu e}$ for normal hierarchy, 
due to matter effect,  
causes the decrease of $P_{\mu\mu}$.}
\label{7k}
\end{center}
\end{figure}

Being interested in maximizing the difference between normal and inverted hierarchy, 
we consider the case when a {\em local maximum} of 
$P_{\mu \mu}$ is  decreased as much as possible.  
This happens when the first minimum 
of $P_{\mu \tau}$, that drives the maximum of $P_{\mu\mu}$,
falls close to the energy where $P_{\mu e}$ is large. 
Such a condition can be  easily analyzed 
numerically, and we use the FORTRAN code  developed for the study \cite{gb} where it is described. 
This code integrates numerically the full three flavor hamiltonian for neutrino 
oscillations in the Earth.  The density of the Earth
is described by the PREM model \cite{prem} and the code has a numerical accuracy  of better than 1 part per million. Moreover, the code has been made publicly available in the 
web resource, 
\begin{quote}
{\tt
http://pcbat1.mi.infn.it/$\sim$battist/cgi-bin/oscil/index.r}
\end{quote}
Let us adopt the result of the 
global analysis 
 of the Bari group \cite{lisi2012}, 
namely, 
$\theta_{12}=33.6^\circ$, 
$\theta_{13}=8.9/9.0^\circ$, 
$\theta_{23}=38.4/38.8^\circ$,
$\delta=194/196^\circ$,
$\Delta m^2_{12}=7.54\times 10^{-5}$ eV$^2$,
$\Delta m^2_{23}=2.39/2.47\times 10^{-3}$ eV$^2$;
where the two values apply to normal/inverted 
hierarchy, respectively.
For normal hierarchy we find that, when the neutrino energy $E$
and the corresponding oscillation length $L$ are  about  
\begin{equation}\label{dirs}
E=6,7,8\mbox{ GeV} \mbox{ and } L=6000,7000,8000\mbox{ km} \mbox{ respectively}
\end{equation}
the maximum of $P_{\mu\mu}$ 
in which we are interested  
lowers to
\begin{equation}P^{max}_{\mu\mu}(\mathrm{NH})\sim 0.7\end{equation}
The difference with $P_{\mu\mu}^{max}(\mathrm{IH})\sim 1$ (i.e., the case of  
inverted hierarchy) is as large as 30\% for the distances $\sim R_\oplus$
of Eq.~\ref{dirs}. 
As an example, we plot in Fig.~\ref{7k} the case $L=7000$ km.

We  verified that for other distances and/or for other energies, the effect diminishes, especially 
when $L$ decreases. Thus Eq.~\ref{dirs} identifies the optimal distances and energies to
search for matter effect on $P_{\mu\mu}$, and impose conditions on the 
type of muon neutrino source and muon detectors needed for this purpose.
We checked that these results do not depend much on the errors on most of the parameters, 
and in particular on CP-violating phase. 
For instance, consider the variation due to $\theta_{23}$; 
the present 3$\sigma$ range  \cite{lisi2012} implies that at 7800 km
$P^{max}_{\mu\mu}(\mathrm{NH})\sim 0.7_{+0.04}^{-0.15}$. This is not a 
very wide variation, and there are reasonable perspectives to improve $\theta_{23}$ in the next few years.  
Note however that if $\theta_{23}$ was larger than currently estimated, the 
effect could increase significantly. 
{More discussion will be given later, in Sect.~\ref{sbl}.}
%

{Appendix \ref{mme} offers some insight into the issue, 
thanks to suitable approximations and some simplified analytical descriptions of  the matter effect. It also collects various relevant formulae; e.g.,  Eqs.~\ref{aq} and \ref{3e2} explain why the effect increases with $\sin^2\theta_{23}$.}


\section{Cross section for neutrino-muon production\label{zsec}}
\begin{figure}[tbp]
\begin{center}
\centerline{\includegraphics[width=7.3cm]{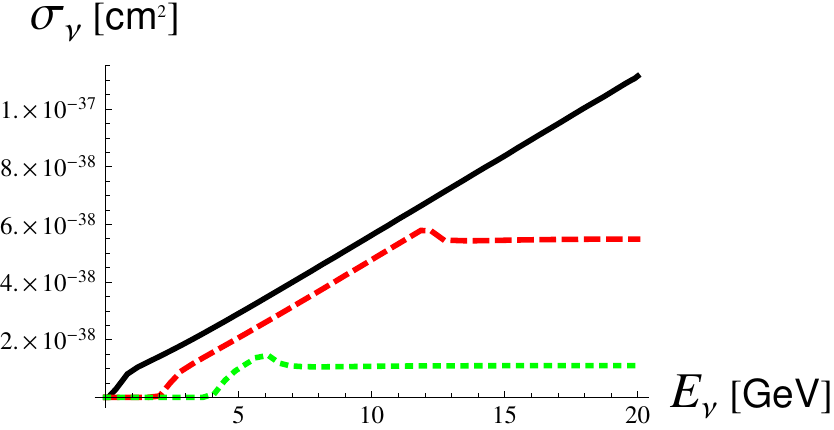}
\includegraphics[width=7.3cm]{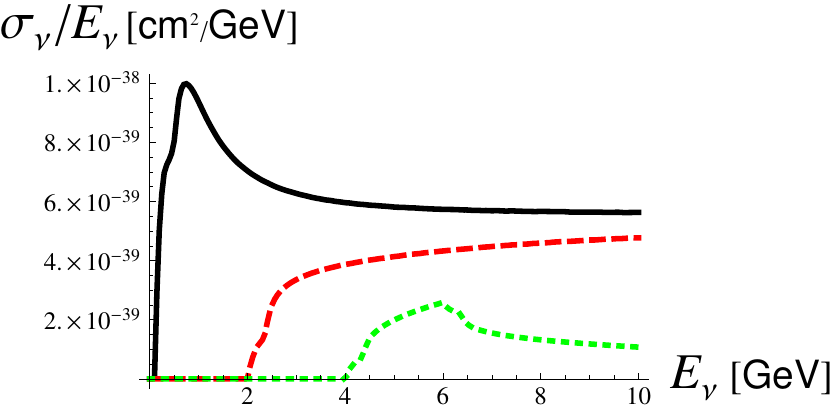}}
\caption{\footnotesize Various 
 $\nu_\mu N\to \mu X$ cross sections  
 as a function of the energy of the incoming $\nu_\mu$. 
The continuous (black) line is the total cross section; the dashed (red) line 
restricts the muon energy to  
$2<E_\mu<12$ GeV;
the dotted (green) to $4<E_\mu<6$ GeV.}
\label{xsec}
\end{center}
\end{figure}

For the energies of interest, the leading interaction that yields observable muons is, 
\begin{equation}
\nu_\mu + N \to \mu^- + X 
\end{equation} 
where $N$ is an average nucleon and $X$ is a set of hadrons; 
we consider water nuclei, therefore a neutron/proton ratio of 4/5. 
We use the differential cross section $d\sigma/dE_\mu$ 
calculated in \cite{shashi}, following \cite{lls}. In view of the fact that the neutrino energies are not very large, we sum the following three contributions
according to the hadronic mass of the final state $m^2_X\equiv p_X^2$:
\begin{enumerate}
\item The quasi-elastic contribution, with $M_A^2=0.95$ GeV$^2$ and $F_A(0)=-1.26$ \cite{ll}. 
\item The delta resonance, where we resort to CVC and PCAC and the parameterization 
of the form factors described in eqs. 12, 13 and 18 of \cite{alv}. 
\item The deep inelastic contribution for $m_X>1.4$ GeV, with GRV94 partons \cite{grv}.
\end{enumerate}
We show in Fig.~\ref{xsec}  the following integrated cross sections:
1)~the total cross section; 
2)~the one where we integrate only the region of muon energy $2<E_\mu<12$ GeV;
3)~the same, restricted to the range $4<E_\mu<6$ GeV.
As customary after \cite{lls}, also the ratio between the cross section and the neutrino energy is 
displayed.\footnote{ 
At the energies in which we are interested, we have found an approximate scaling of the integrated cross section: 
$\sigma(E_\nu, E_1<E_\mu<E_2)\equiv 
\int_{E_1}^{E_2}  \frac{d\sigma}{dE_\mu}(E_\nu,E_\mu)\ dE_\mu\approx
f(E_\nu-E_1,E_2-E_1)$.} 
Note that when $E_\nu$ is close to the threshold, the contribution of individual 
resonances considered (the nucleon and the delta) is clearly visible.

\section{Needs for the experiment\label{rfe}}
\subsection{Mass of the detector\label{r1}}
The number of muon events 
scales with the mass of the detector $M_{\mbox{\tiny det}}$ and the distance $L$ as 
$N_{\mu}\propto M_{\mbox{\tiny det}}/L^2$. Thus, in order to have 
as many events as those in a detector of 10 kton at 800 km from the source (roughly 
corresponding to the present generation long baseline experiment NO$\nu$A \cite{nova}) 
 at a distance of 8000 km, we need a detector of about 1 Mton. 
Moreover, keeping in mind Fig.~\ref{7k}, it is easy to be convinced 
that we need to detect muons from few to ten GeV,  
which implies that we have to measure tracks of $10-40$ meters in water. 

These requirements, concerning the mass of the detector 
and the length of the tracks to be revealed, point toward a  
large but relatively simple underwater or under-ice detector, that could resemble the PINGU \cite{pingu}
and ORCA \cite{orca} proposals that are being developed/considered by the IceCube and KM3NET collaborations, respectively. In the following, we will assume that the detector has 
a number of useful target nucleons of 
\begin{equation}\label{trg}
N_{\mbox{\tiny targ.}}=6\times 10^{35}
\end{equation}
corresponding to a cube of water of 100 m in size, i.e., $M_{\mbox{\tiny det}}=1$ Mton, 
and leave a more detailed description of the detector for future work. 
Our results can be simply rescaled with the mass of the detector. 

\subsection{Source-detector distance\label{r2}}
The next question is which arrangement of source and detector would fit the optimal range of 
distances identified above. Let us consider the existing neutrino lines \cite{pdg}
and a few sites of under-water and under-ice neutrino experiments.
Their approximate coordinates in degrees are
%
%
\begin{equation}
(\lambda,\phi)=\left\{
\begin{array}{cl}
( +41.8, -88.3)& \mbox{Fermilab}  \\
( +46.2,+6.0 )& \mbox{CERN}  \\
(+36.4 ,+140.6 )& \mbox{J-PARC} 
\end{array}\right. \mbox{ and } 
=\left\{
\begin{array}{cl}
(-90 , +0.0 )& \mbox{South Pole}  \\
( +36.3,+16.1 )& \mbox{Sicily}  \\  
( +51.8,+104.3 )& \mbox{Baikal Lake} 
\end{array}\right. 
\end{equation}
`South Pole', `Sicily'  and `Baikal Lake'
correspond to the coordinates  of IceCube, KM3NET and GVD, respectively.
From the position versors 
$\vec{n}_i=(\cos\lambda_i\cos\phi_i,\cos\lambda_i\sin\phi_i, \sin\lambda_i)$
we find the relative distances as
$L=R_\oplus  |\vec{n}_1-\vec{n}_2|=R_\oplus \sqrt{2 (1-\vec{n}_1\cdot \vec{n}_2)}$;
or, using the web resource \cite{daft} one gets the 
minimum distance on the Earth surface $A$ (the arc) and thus 
$L=2 R_{\oplus} \sin[A/(2 R_{\oplus})]$. We find the following 
distances, expressed in km

\centerline{
\begin{tabular}{c|ccc}
                  &       Fermilab & CERN & J-PARC \\ \hline
South Pole &      11600  & 11800  &  11400 \\
 Sicily          &        {\bf 7800}  & 1230 &   9100 \\
Baikal Lake &       8700  &  {\bf 6300} &  3300 
\end{tabular}
}

\begin{figure}[tbp]
\begin{center}
\centerline{\includegraphics[width=8cm]{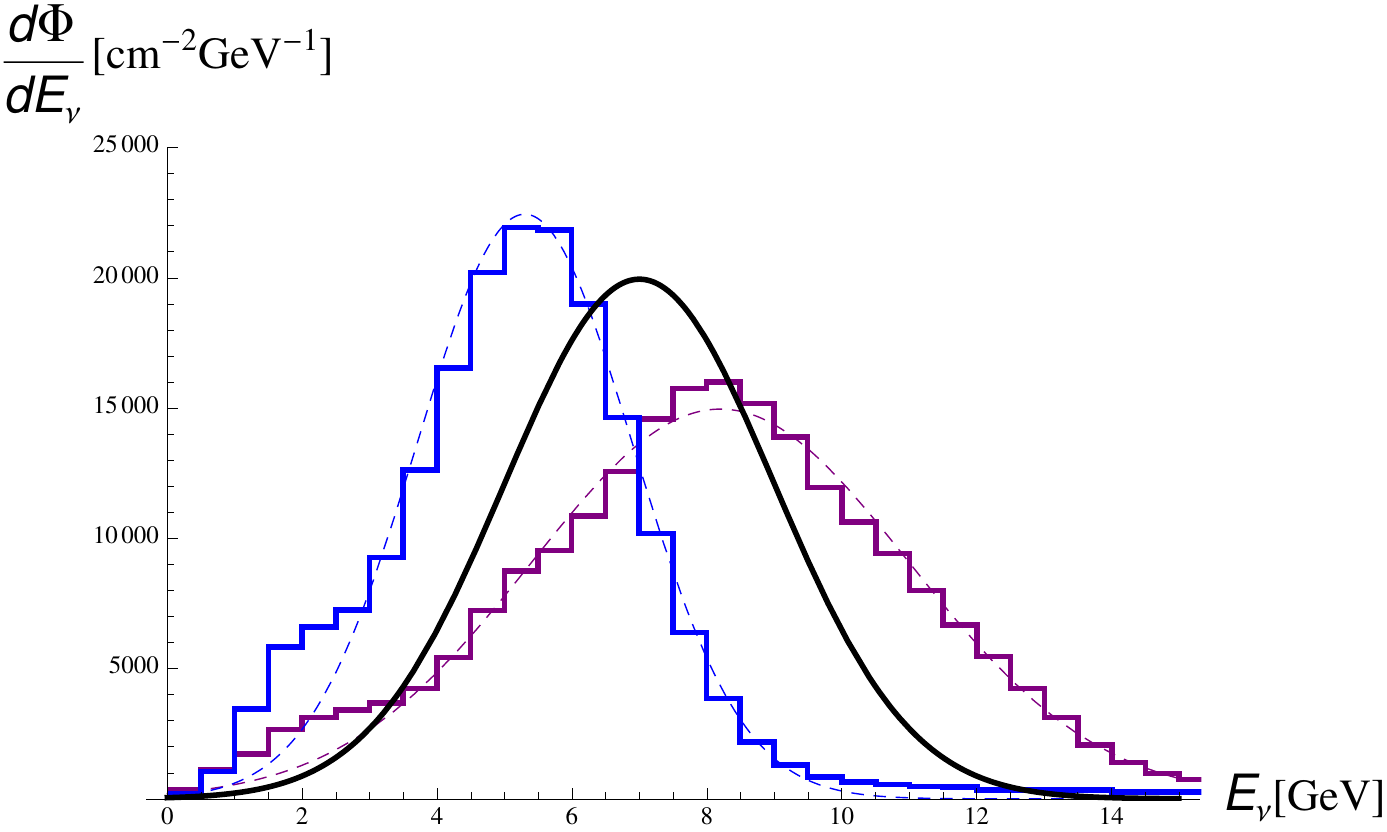}}
\caption{\footnotesize Differential neutrino fluences for a distance of $L=7800$ km. 
The two dashed gaussian distributions at the right and at the left are quite similar to the 
fluxes ME and HE of NuMI beam \cite{kopp} for $10^{20}$ protons-on-target, 
shown as histograms. The black gaussian distribution with intermediate energies
is assumed for the subsequent calculations. Compare with Fig.~\ref{7k}.}
\label{fluer}
\end{center}
\end{figure}

\vskip1mm
\noindent The pairs CERN to Baikal Lake and  
Fermilab to Sicily are at optimal distances. 
The first case would be somewhat more convenient, for its  
$1/L^2$ factor is 50\% larger. 
However, as we will show, also the second case offers reasonable 
opportunities for an experiment, and 
before continuing, we note that also the location of ANTARES,  with 
$(\lambda,\phi)=(+42.8, +6.2)$, would be also  appropriate, its distance from Fermilab being 6900 km.

\subsection{Properties of the neutrino beam\label{r3}}
The existing neutrino beam from CERN has  a mean energy twice
higher than necessary to probe the matter effect, whereas it has already been shown that the NuMI neutrino beam 
of Fermilab can be arranged to fit our needs  \cite{kopp}.  
The estimated fluences, and more precisely those called HE and ME options \cite{kopp},  
are evaluated for $10^{20}$ protons on target and at 1.04 km from the source.
They  can be reasonably approximated as gaussian distributions,
proportional to $G(E,\bar{E},\delta E)=\exp[-(E-\bar{E})^2/(2 \delta E^2)]/\sqrt{2\pi\ \delta E^2}$, with 
$\bar{E}=8.2$ (resp., 5.3) GeV and $\delta E=2.8$ (resp., 1.6) GeV as shown in 
Fig.~\ref{fluer}, where we  rescaled for a distance of $L=7800 \mbox{ km}$.
The fluence of a conventional muon neutrino beam, 
with properties intermediate between the above two cases 
(see Fig.~\ref{fluer}) is approximated by,
\begin{equation} \label{fro}
\frac{d\Phi}{dE}=\frac{ 10^{5}\ \nu_\mu}{\mbox{cm}^2} \times G(E,\bar{E},\delta E)
\ \ \ \mbox{ with } \ \ \ 
\left\{
\begin{array}{c}
\bar{E}=7\mbox{ GeV}\\[1ex]
\delta E=2\mbox{ GeV}
\end{array}
\right.
\end{equation}
It corresponds to a beam of $10^{20}$ protons on target sent at a distance of 7800 km.
In the following, we will consider such a fluence for definiteness.

\section{Results}
\begin{figure}[tbp]
\begin{center}
\centerline{\includegraphics[width=8.2cm]{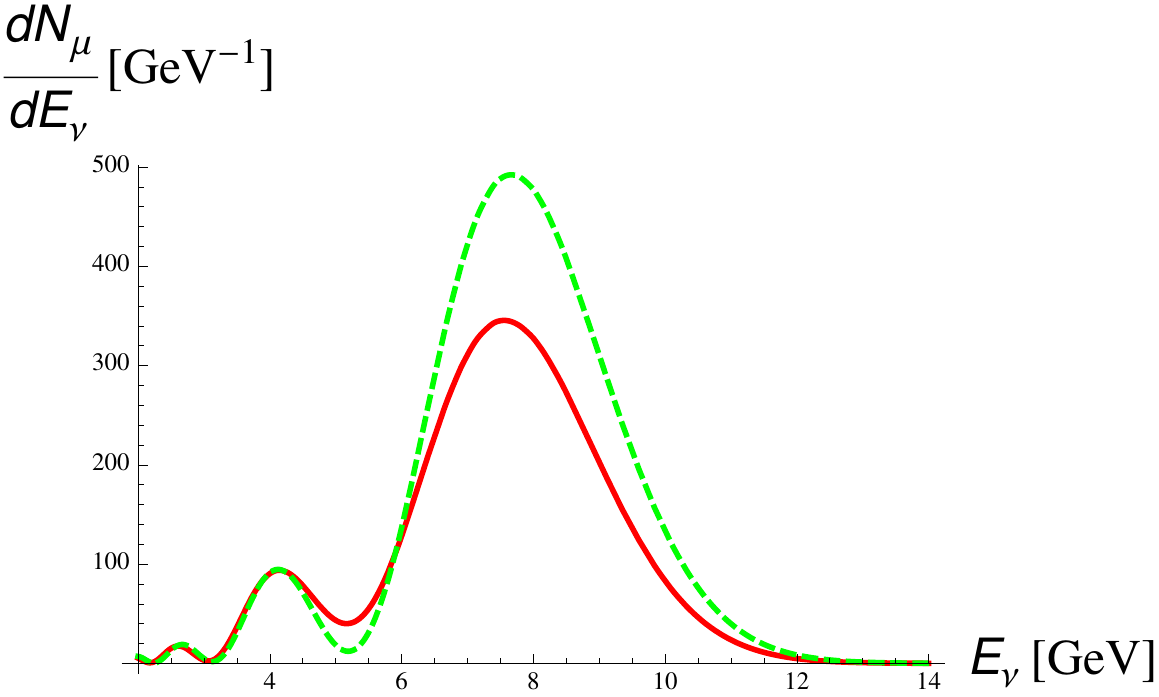}
\includegraphics[width=6.2cm]{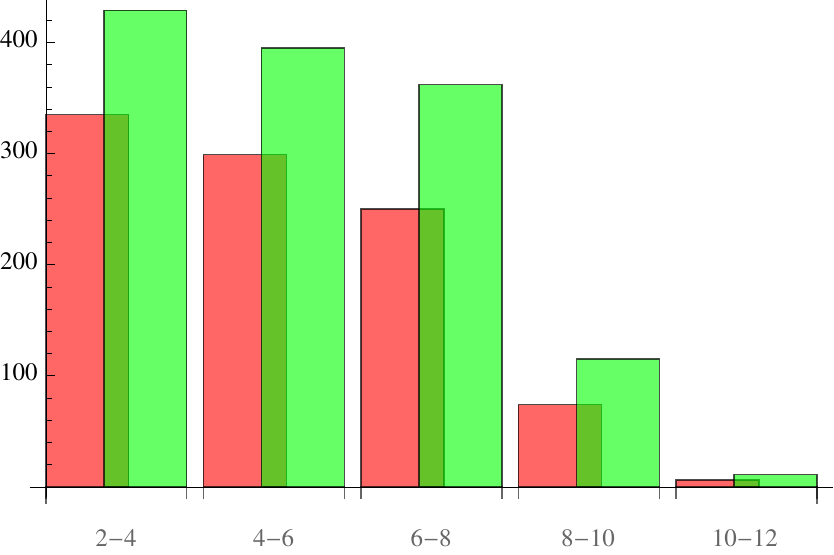}}
\caption{\footnotesize Left panel, distribution in the neutrino energy; 
the continuous curve (red) is for normal hierarchy, the 
dashed one (green) is for inverted hierarchy. Right panel,  
number of events 
in various windows of muon energy (2-4 GeV, 4-6 GeV, etc., as indicated below the 
abscissa)  
for normal hierarchy (left bars in the back, red) and for inverted hierarchy 
(right bars in front, green).}
\label{ddd}
\end{center}
\end{figure}

The number of expected muon tracks contained in the detector, corresponding to 
muon energies larger than $E_{th}$ and to a given fluence of muon neutrinos $d\Phi/dE$, is 
\begin{equation}
N_{\mu}=
N_{\mbox{\tiny targ.}}\times \int_{E_{th}}  P_{\mu\mu}(E)\times \frac{d\Phi}{dE}(E)\times \sigma(E)\ dE
\end{equation}
When we use the number of targets as in Eq.~\ref{trg}, the 
fluence of Eq.~\ref{fro} (see previous section)
and the cross section to produce muons above 2 GeV and below 12 GeV
(Sect.~\ref{zsec}) we get
\begin{equation}
N_{\mu}=
\left\{
\begin{array}{l}
950 \mbox{ with normal hierarchy}\\[1ex]
1300 \mbox{ with inverted hierarchy}
\end{array}
\right.
\end{equation}
The difference is 30\%, as expected. The distribution of the events is shown in
Fig.~\ref{ddd}. Even if we retain only the events with muon energy above 4 GeV, in order to 
use a simpler detector or to avoid confusion with 
atmospheric neutrinos, we loose only 1/3 of the events.
The number of signals is ten times the expected statistical variance, 
$\sqrt{N_\mu}$, so that the difference between normal and inverted hierarchy 
cannot be mimicked by  any plausible statistical fluctuation. 
The signal events can be distinguished by those from atmospheric neutrinos not only because of their spectra, but most of all, because  the neutrinos arrive in bunches, due to the pulsed structure of the artificial neutrino beam, and to some extent also thanks to a certain degree of directionality at these energies--the angle between the neutrino and the produced muon being 
$\sim\sqrt{\mbox{1 GeV}/E}\sim $ several tens of degrees. 

It is also possible to cross check the results by using 
a muon anti neutrino beam (from $\pi^-$ decay). In fact,  
the matter effect  works in the opposite manner on antineutrinos,  
diminishing by 30\% the number of $N_\mu$ for inverted hierarchy, rather 
than for normal hierarchy as for the neutrino beam considered above.

{
\subsection{Stability of the results\label{sbl}}
In principle, various effects could affect the expectations, but none among those that we have considered is larger than  few percent:\\
1) If the true oscillation parameters are not the present best-fit values, the number of expected muons will change.  
  We find
$N_\mu(\mathrm{NH})^{-3.5-3.0+0.9\%}_{+3.4+2.7-2.1\%}$  
and 
$N_\mu(\mathrm{IH})
{}^{-1.3+0.0-0.4\%}_{+1.0+0.0-0.3\%}$, 
where  the superscript (resp., the subscript) percentage  variations 
correspond to the upper (resp., lower) 1$\sigma$ value 
of  $\theta_{23}$, $\theta_{13}$ and  $\Delta m^2_{23}$ given in \cite{lisi2012}; the variations with the 
 other parameters are (much) smaller.\footnote{{
 The smallness of the variation due to  $\Delta m^2_{23}$ is due to the selected neutrino beam: indeed, the signal has been almost maximized with the present best-fit values of the oscillation parameters.}} 
Thus, the expected changes are small. Note also that our knowledge of the oscillation parameters is likely to improve in the near future, diminishing this uncertainty further. \\
2) What would happen if some electron event contaminated the muon sample? 
In the range of energy between 2 and 12 GeV, 
we expect 390 electron-type events for normal hierarchy and less than 10 electron-type events for inverted hierarchy. Thus, it is enough that we reject 90\% or more of this type of events, in order to have a contamination of the muon sample of 4\% or less, that does not affect the result. (Note, incidentally, that if an electron sample can be clearly identified, this will be an additional signal that the matter effect is occurring on neutrinos. However, the electron signal depends quite crucially on the specific type of detector, and we prefer to stress the muon signal in this discussion, in view of its relative easiness of detection.).\\
3) Finally, we discuss the muon events caused by charged current interactions of tau neutrinos,  followed by the decay $\tau\to \mu+ \bar{\nu}_\mu +\nu_\tau$. The probability of oscillation $P_{\mu\tau}$ is large, but the beam is not optimized to see taus;  moreover,  the charged current cross section has a relatively high energy threshold. Due to these circumstances, and to the branching ratio BR$=17.41\%$, 
this reaction yields only 26 (resp., 24) muons for normal (resp., inverted) hierarchy 
with energies between 2 and 12 GeV.\footnote{{The non-optimal beam energy gives a factor $\sim 1/2$; the tau-production threshold, along with the fact that part of the energy is carried away by the two neutrinos in the final state, gives a factor $\sim 1/3$. These factors, along with the BR, 
explain the difference with the $\sim 1000$ muon signal discussed above.}}   
We conclude that the ensuing number of muon events is small. Moreover, the difference due to these additional muon events between the two hierarchies is negligible.}

\section{Summary and Discussion}
The existing experiments are compatible with two 
 possible neutrino mass spectra: the 
 normal hierarchy  
might resemble the spectra of the other fermions, depending on the mass
of the lightest neutrino, but the inverted hierarchy surely does not.
Presumably, a 
proof of inverted hierarchy would be as shocking and informative as the discovery of a sterile neutrino.  
The study of the matter effect 
in the Earth is the most direct way to identify  the neutrino mass spectrum.
Recall that the neutrino spectrum should be known to disentangle the effect of 
leptonic CP violation, and thus to measure the CP violating phase; moreover, the spectrum matters also 
for the interpretation of neutrinoless double beta decay, of cosmological search for neutrino masses, of supernova neutrinos, etc.

In this work, {we have discussed a long baseline neutrino experiment, aimed to 
study the matter effect in the Earth, by taking advantage of a large muon disappearance.}
We have argued that the ideal conditions  
are obtained for muon neutrinos with average energies 
of $6$ (8) GeV, that propagate for a distance of  $L=6000$ (8000) km. 
A previous similar proposal, \cite{dick}, discussed larger detectors with higher energy
thresholds modeled on neutrino telescopes; 
other apparently related proposals \cite{wint,sp}  considered
electron neutrino detection instead, that is significantly 
more demanding than muon detection.

We have shown 
that a conventional 
(water Cherenkov) muon detector of 1 Mton 
along with a conventional muon neutrino beam ($10^{20}$ protons on target) 
can observe $\sim$1000 signal events, even  
 at a relatively large distance of 7800 km. 
The signal is composed by muons between $2-10$ GeV, 
well characterized experimentally, and  the muons   
can be identified as $10-40$ m long tracks;
moreover, 2/3 of them are above 4 GeV. 
Due to the matter effect, the inverted hierarchy yields 30\% more events than the normal hierarchy case.
Thus, the difference induced by the matter effect
is quite large and even a moderate understanding of the artificial neutrino 
beam should suffice to identify experimentally which is the neutrino mass spectrum. 


\section*{Acknowledgments}
We
thank 
G.~Battistoni,
A.~Capone,  
R.~Coniglione, 
P.~Coyle, 
G.V.~Domogatsky, 
S.~Galat\`a, 
M.~Goodman, 
P.~Lipari, 
S.~Ragazzi,
G.~Riccobene,
F.~Terranova, 
A.~Varaschin,
L.~Votano
{and an anonymous Referee of EPJC
for useful discussions. 
FV is grateful to the Organizers and the Participants in the Orca meeting at Paris 
for precious feedback \cite{paris}.}

\appendix\section{Remarks on the matter effect\label{mme}}
In this appendix, we examine the oscillations by employing some simplifying assumptions, in order to 
obtain a qualitative understanding of the results in Sect.~\ref{mp}: we consider  oscillations with a single scale, and we also consider oscillations in constant matter density.
In fact, the second hypothesis  is rather inaccurate in the conditions in which we are interested, and 
we need (and we use) a more accurate evaluation of the oscillation probabilities 
for the actual calculations. However, a qualitative discussion based on simple-minded analytical results complements usefully the numerical results discussed in the main text.

We try to go immediately to the main point, postponing derivations and refinements.
Under suitable assumptions
the probability that a muon converts into an electron 
is simply,  
\begin{equation}
P_{\mu e}=\sin^2\theta_{23}\ \sin^22\widetilde{\theta_{13}} \ \sin^2\widetilde{\varphi} 
\mbox{ with }\widetilde{\varphi} =\frac{\widetilde{\Delta m^2} L}{4 E}  \label{aq}
\end{equation}
Most of the results in which we are interested
follow from the above simple formula.  
In Eq.~\ref{aq}, we introduced the usual matter-modified mixing angle and squared-mass-difference 
\begin{equation} \label{peta}
\left\{
\begin{array}{l}
\sin2\widetilde{\theta_{13}}=\sin2{\theta_{13}}\ / \Delta \\[.3ex]
\cos2\widetilde{\theta_{13}}=(\cos2{\theta_{13}} -\varepsilon) / \Delta \\[.3ex]
\widetilde{\Delta m^2}={\Delta m^2}\times \Delta 
\end{array}
\right.
\mbox{ where }
\Delta=\pm \sqrt{ (\cos2\theta_{13}-\varepsilon)^2 + \sin^22\theta_{13}} 
\end{equation}
The sign of $\Delta$ is  
matter of convention;
the ratio between matter and vacuum term is,
\begin{equation}\label{msw}
\varepsilon\equiv \pm \frac{\sqrt{2 } G_F n_e }{\Delta m^2/(2 E)}
\approx \pm \frac{\rho}{5.5 \mbox{ g/cm$^3$}}\times
\frac{Y_e}{1/2}\times 
\frac{2.4\times 10^{-3}\mbox{eV}^2}{\Delta m^2}\times
\frac{E}{5.5 \mbox{ GeV}}
\end{equation}
where $G_F$ is the Fermi coupling 
and we identify $\Delta m^2$ with $\Delta m^2_{23}$.
Now, instead, the sign is important: it is plus for normal hierarchy and minus for inverted hierarchy.
Considering the average matter density of the Earth
$\rho=5.5 $ g/cm$^3$ and $Y_e=1/2$, we get 
$n_e= 1.7\times 10^{24}$ e$^-$/cm$^3$ for  the electronic density.
Thus, the 
characteristic length of MSW theory is,
\begin{equation}
L_* \equiv \frac{1}{\sqrt{2} G_F n_e}\sim 1000\mbox{ km}
\end{equation}
We see that, for normal hierarchy, the maximum of $P_{\mu e}$ 
 obtains when: $(1)$ $\Delta $ is as small as possible, in order to maximize 
 $\sin2\widetilde{\theta_{13}}$; moreover, 
$(2)$ the phase of propagation is $\widetilde{\varphi} \sim \pi/2$. These conditions are met when the neutrino energy and the propagation distance are,
\begin{equation}
\label{masul}
E_{\mathrm{\tiny max}}=\frac{\Delta m^2 L_*}{2}\cos2\theta_{13}\sim 5.5\mbox{ GeV} \mbox{ and }
L_{\mathrm{\tiny max}}=\frac{\pi L_*}{\tan2\theta_{13}}\sim 9000\mbox{ km}
\end{equation}
In the case of inverted hierarchy, the matter effect depresses $P_{\mu e}$, that becomes negligible.

In principle, one could check this simple prediction concerning 
$P_{\mu e}$, 
however it is practically easier to study muons rather
than electrons. Then, let us consider the  survival probability $P_{\mu \mu}$, focussing again
on the normal hierarchy case.  
We want  that a local maximum of $P_{\mu \mu}$, resulting 
from $P_{\mu \tau}$ and from $P_{\mu e}$, is as small as possible.
Thus, we are interested in the case when the minimum of $P_{\mu \tau}$ happens in the vicinity of the energy identified in Eq.~\ref{masul}. When the phase of oscillation of $P_{\mu\tau}$ is close to the vacuum phase, the condition $\Delta m^2 L/(2 E_{\mathrm{\tiny max}})=2\pi$ gives  $L\sim 6000$ km. This suggests that the distance that amplifies the matter effect on $P_{\mu\mu}$ is between 6000 and 9000 km, that does not disagree severely with the quantitative conclusions of the precise 
numerical analysis of Sect.~\ref{mp}.

Finally, we collect more arguments and technical remarks concerning the matter effect.
Let us write in full generality 
 the amplitude of three flavor neutrino oscillations
\begin{equation}
\mathcal{A}=\mbox{Texp}\left[-i\int dt \mathcal{H}_\nu(t)\right]=R_{23} R_\delta \left(
\begin{array}{ccc}
a_{11} & a_{12} & a_{13} \\
a_{21} & a_{22} & a_{23} \\
a_{31} & a_{32} & a_{33} 
\end{array}
\right) R_\delta^* R_{23}^t \label{cagon}
\end{equation}
where $a_{ij}$ depend upon $\Delta m^2_{23}$, $\theta_{13}$,
 $\Delta m^2_{12}$, $\theta_{12}$, and $H=\pm 1$ (the type of mass hierarchy,
 normal/inverted): 
see in particular Eqs.\ 1, 3, 5, 6  of \cite{gb}. By solving numerically the 
evolution equations, we calculate the 
complex numbers $a_{ij}$ and therefore the amplitudes and the probabilities.
At this level, there is no approximation (except the numerical ones).

When the ``solar'' $\Delta m^2_{12}$ is set to zero--i.e., when its effects
 are negligible--the only non-zero out-of-diagonal elements $a_{ij}$ in Eq.~\ref{cagon}
 are $a_{13}$ and $a_{31}$.
The CP violating phase $\delta$ drops out from the 
probabilities $P_{\ell\ell'}=|\mathcal{A}_{\ell'\ell}|^2$, that moreover  
become symmetric, $P_{\ell\ell'}=P_{\ell'\ell}$ for each $\ell,\ell'=e,\mu,\tau$.
Therefore, in this approximation 
we have 3 independent probabilities and all the other ones are fixed. 
We can chose, e.g., 
\begin{equation} \label{3e}
P_{e\mu}=\sin^2\theta_{23} |a_{13}|^2,\ 
P_{e\tau}=\cos^2\theta_{23} |a_{13}|^2, \ 
P_{\mu\tau}=\sin^2 \theta_{23} \cos^2 \theta_{23} |a_{33}-a_{22}|^2, 
\end{equation}
so that, e.g., $P_{ee}=1- P_{\mu e}-P_{\tau e}=|a_{11}|^2$. From these formulae we obtain
 \begin{equation} \label{3eb}
P_{\mu e}=\sin^2\theta_{23}\ (1-P_{ee})\mbox{ and }
P_{\mu\tau}=\frac{1}{4}\sin^2 2 \theta_{23} \left|1-\sqrt{P_{ee}}\ e^{i\hat\varphi}\right|^2, 
\end{equation}
where $\hat\varphi$ is a (rapidly varying) phase factor. 
Two important remarks are in order: 
 \begin{enumerate}
\item The last equation shows that  $P_{\mu e}$ is large in the region 
where $P_{ee}$ is small, and that $P_{\mu\tau}$ remains close to zero 
in the first non-trivial minimum near $\hat\varphi=2\pi$, even when $P_{ee}\approx 0.3-0.4$ due to matter effect. 
\item The sign of 
$\Delta m^2$ controls the sign of the vacuum hamiltonian; therefore, switching between the two 
mass hierarchies or switching between neutrinos and antineutrinos has the same effect; e.g.,
$P_{{e}{\mu}}(\mbox{IH})=P_{\bar{e}\bar{\mu}}(\mbox{NH})$.
\end{enumerate}
The first remark is consistent with our numerical findings, that $P_{\mu e}$ is amplified and 
 $P_{\mu\tau}$ does not deviate strongly from its behavior in vacuum 
 in the conditions that are relevant for our discussion.

Proceeding further with the approximations, and considering at this point 
the case of constant matter density, we obtain simple and closed expressions. For the case of  
 normal mass hierarchy, they read: 
 \begin{equation} \label{3e2}
 \begin{array}{l}
 a_{13}=a_{31}=-i \sin\widetilde{\varphi}  \sin2 \widetilde{\theta_{13}} \\[.2ex]
 a_{11}=\cos\widetilde{\varphi} + i \sin\widetilde{\varphi} \cos2 \widetilde{\theta_{13}}=a_{33}^* \\[.5ex]
 a_{22}=\cos\widetilde{\varphi}' + i \sin\widetilde{\varphi}' 
  \end{array}
 \end{equation}
 where 
\begin{equation}\widetilde{\varphi}'=\frac{{\Delta m^2}  L}{4 E} (1+\varepsilon)\end{equation}
From  Eqs.~\ref{3e} and \ref{3e2}, we recover the expression of Eq.~\ref{aq}, used in the above discussion.
%
In the approximation of constant matter density, the phase $\hat\varphi$ entering 
the expression of the probability $P_{\mu\tau}$ 
is given by $\sqrt{P_{ee}}\cos\hat\varphi\equiv \cos\widetilde\varphi \cos\widetilde\varphi'-
 \sin\widetilde\varphi \sin\widetilde\varphi'\cos 2\widetilde{\theta_{13} }$. This is close to 
 the vacuum phase when 
 $\varepsilon$ is large or small in comparison to 1: in fact,  we have
 $\cos 2\widetilde{\theta_{13} }\sim \pm 1$ and  
 $\widetilde\varphi\sim \pm \Delta m^2 L/(4 E) (1-\varepsilon)$ from Eq.~\ref{peta}, 
 so that $\cos\hat\varphi\sim 
 \cos[\Delta m^2 L/(2 E)]$.

\end{document}